{

\font\titlefont=cmr10 scaled\magstep3

\nopagenumbers

\topskip200truept

\centerline{ \titlefont Zeta function regularization }
\centerline{ \titlefont in de Sitter space: }
\centerline{ \titlefont the Minkowski limit }

\vskip50truept

\centerline{
Alan Chodos\footnote{$^1$}
{chodos@yalph2.physics.yale.edu}
and
Andr\'as Kaiser\footnote{$^2$}
{kaiser@hepunix.physics.yale.edu}
}

\centerline{ Center for Theoretical Physics,
Sloane Physics Laboratory }
\centerline{ Yale University, New Haven, Connecticut
06520-8120 }

\vfill

\eject

}

{ \bf Abstract: }

\smallskip

We study an integral representation for the zeta function 
of the one-loop effective potential for a minimally coupled 
massive scalar field in D-dimensional de Sitter spacetime. 
By deforming the contour of integration we present it in a 
form suitable for letting the de Sitter radius tend to infinity, 
and we demonstrate explicitly for the case D=2 that the 
quantities $\zeta( 0 )$ and $\zeta'( 0 )$ have the appropriate 
Minkowski limits. 

\bigskip

{ \bf I. Introduction }

\smallskip

In this paper we examine some properties of the effective 
potential for a minimally coupled scalar field in de Sitter 
spacetime. We regard this as a preliminary study with application 
to symmetry breaking effects in de Sitter space, which can have 
significant cosmological consequences [1]. These can be 
investigated if one also includes a potential for the scalar field. 

Our particular concern in this work is to examine the regularization 
of the effective potential by the zeta-function method [2], and to 
show that after such regularization one can compute explicitly 
the limit of the potential as the de Sitter radius $\alpha$ tends 
to infinity, and recover thereby the Minkowski space result. 

Of course de Sitter space itself becomes Minkowski space as 
the radius becomes infinite, and furthermore at the naive level 
the limit of the de Sitter zeta function $\zeta( s )$ goes over 
into the Minkowski one (see below). However in order to regularize 
the effective potential one must perform an analytic continuation 
of $\zeta$ to a neighborhood of $s = 0$. After this this is done, 
it is no longer evident (or even necessarily true) that the 
zeta-function possesses the correct $\alpha \rightarrow \infty$ 
limit. 

In other words, the process of analytic continuation and the 
$\alpha \rightarrow \infty$ limit may not commute with each other, 
and it is reassuring to check explicitly that the analytically-continued 
$\zeta( s )$ still possesses the expected Minkowski limit. 

A simple example of what needs to be checked is provided by the 
series 

$$
f( s ; a ) = 
\sum_{ n = 0 }^{ \infty } 
e^{ -( n + 1 ) \alpha ( s - 2 ) } 
( -1 )^n 
\eqno(1.1)
$$

($\alpha > 0$), which for $\Re s > 2$ clearly obeys 

$$
\lim_{ \alpha \rightarrow \infty } 
f( s ; a ) = 0 
\eqno(1.2)
$$

But $f( s ; a )$ possesses an analytic continuation  in $s$ 

$$
f( s ; a ) = 
{ 1 \over 
1 + e^{ \alpha ( s - 2 ) } 
} 
\eqno(1.3)
$$

which, in a neighborhood of $s = 0$ has the limit 

$$
\lim_{ \alpha \rightarrow \infty } 
f( s ; a ) = 1 
\eqno(1.4)
$$

Let's take a look at the Minkowski zeta-function first. 
In $D$ dimensions, it is defined by 

$$
\zeta( s )
\ = \
\sum_k
{ \mu^{ 2 s } \over ( k^2 + m^2 )^s }
\eqno(1.5)
$$

where $\mu$ is a scale factor with dimensions of mass, 
that is included to make the dimension of $\zeta$ independent 
of $s$.

In Minkowski spacetime the sum is really continuous, 
so we can replace $\sum_k$ with $V d^D k / ( 2 \pi )^D$ 
($V$ is the volume of the spacetime) to transform it 
into an integral form 

$$
\zeta( s )
\ = \
V \int
{ d^D k \over ( 2 \pi )^D } \
{ \mu^{ 2 s } \over ( k^2 + m^2 )^s }
\eqno(1.6)
$$

or divide by $V$ to obtain the $\zeta$ 
density 

$$
\zeta( s )
\ = \
\int
{ d^D k \over ( 2 \pi )^D } \
{ \mu^{ 2 s } \over ( k^2 + m^2 )^s } \ \ \ \ . 
\eqno(1.7)
$$

\bigskip

{ \bf II. de Sitter zeta-function }

\smallskip

Proceeding to the de Sitter case we note that D-dimensional 
de Sitter spacetime can be embedded in a D+1 dimensional 
Minkowski-space (we use signature $<+,-,-,...,->$), where it 
is a D dimensional hypersurface determined by 

$$
z^{_02} - \sum_i z^{_i2} 
\ = \ 
- \alpha^2 
\eqno(2.1)
$$

We use the following parametrization of this space 

$$
\eqalign{ 
z_0 &= \sinh( x_0 ) \cr
z_1 &= \cosh( x_0 ) \cos( x_1 ) \cr
z_2 &= \cosh( x_0 ) \sin( x_1 ) \cos( x_2 ) \cr
&... \cr
} 
\eqno(2.2)
$$

We start with the de Sitter action 

$$
S 
\ = \ 
\int d^D x \ 
\bigg\{ \ 
\partial_\mu \phi^* 
\partial^\mu \phi 
\ - \ 
m^2 | \phi |^2 \ 
\bigg\} 
\eqno(2.3)
$$

Substituting $x_0 = - i x_4$, we obtain the Euclidean action 

$$
S 
\ = \ 
\int d^D x \ 
\bigg\{ \ 
\partial_\mu \phi^* 
\partial^\mu \phi 
\ + \ 
m^2 | \phi |^2 \ 
\bigg\} 
\eqno(2.4)
$$

The classical $\phi( x )$ field configurations form a 
vector space with the scalar product 

$$
< \phi , \psi > \ \ = \ 
\int d x \ \phi^*( x ) \ 
\psi( x ) 
\eqno(2.5)
$$

and the second variation of the action 

$$
{ \delta^2 S \over \delta \phi( x ) \delta \phi^*( y ) } 
\ = \ 
( \sqcap \mkern-16mu \sqcup_E 
+ m^2 ) \ 
\delta( x - y ) 
\eqno(2.6)
$$

is similarly considered a matrix (bilinear operator 
on the vector space). The zeta function is defined 
with the eigenvalues of this matrix 

$$
\zeta( s ) \ = \ 
\sum_i { \mu^{ 2 s } \over \lambda_i^s } 
\eqno(2.7)
$$

where $\mu^{ 2 s }$ is included to get 
a dimensionless quantity. 
When we substitute $x_0 = - i x_4$, 
$\cosh( x_0 )$ transforms into $\cos( x_4 )$, 
and $\sinh( x_0 )$ into $- i \sin( x_4 )$. 
This means that now we have 

$$
\eqalign{
z_0 \ &= \ - i \, \sin( x_4 ) \cr
z_1 \ &= \ \cos( x_4 ) \, \cos( x_1 ) \cr
z_2 \ &= \ \cos( x_4 ) \, \sin( x_1 ) \, \cos( x_2 ) \cr
&... \cr
}
\eqno(2.8)
$$

which shows that Euclidean de Sitter space is a 
(series of) D-sphere(s), therefore the eigenfunctions 

(eigenvectors) of $\delta^2 S / 
\delta \phi( x ) \delta \phi( y )$ are generalized 
spherical functions, with the eigenvalue 

$\lambda = l ( l + D - 1 ) / \alpha^2 \ + \ m^2$. 
The multiplicity of the modes with the same $l$ is [3] 

$$
{ \Gamma( D + l + 1 ) \over 
\Gamma( D + 1 ) \Gamma( l + 1) } \ 
- \ 
{ \Gamma( D + l - 1 ) \over 
\Gamma( D + 1 ) \Gamma( l - 1 ) } 
\eqno(2.9)
$$

so 

$$
\zeta( s ) \ = \ 
\sum_l \ 
\Bigg[ \ 
{ \Gamma( D + l + 1 ) \over 
\Gamma( D + 1 ) \Gamma( l + 1 ) } \ 
- \ 
{ \Gamma( D + l - 1 ) \over 
\Gamma( D + 1 ) \Gamma( l - 1 ) } \ 
\Bigg] 
\ 
{ \mu^{ 2 s } \over 
\Big[ 
{ l ( l + D - 1 ) \over \alpha^2 } 
+ m^2 
\Big]^s } 
\eqno(2.10)
$$

$$
\ = \ 
{ \mu^{ 2 s } \alpha^{ 2s } 
\over 
\Gamma( D + 1 ) } \ 
\sum_l \ 
\Big[ \ ( D + l ) ( D + l - 1 ) \ - \ 
l ( l - 1 ) \ \Big] \ 
{ \Gamma( D + l - 1 ) \over \Gamma( l + 1 ) } 
\ 
{ 1 \over 
\big[ \ 
( l + { D - 1 \over 2 } )^2 
\ - \ \beta^2 \ 
\big]^s } 
\eqno(2.11)
$$

where $\beta^2 = - m^2 \alpha^2 + 
\big( { D - 1 \over 2 } \big)^2$. 

Let's make a brief digression here, and check how the 
$\zeta$-density approaches the Minkowski case as 
$\alpha \rightarrow \infty$. Divide $\zeta$ by the volume 
of de Sitter space, $\alpha^D V_D$, where $V_D$ is 
the surface area of a $D + 1$-dimensional sphere, 
in other words the volume of the manifold $S^D$. 

$$
{ 1 \over \alpha^D V_D } 
\zeta( s ) \ = \ 
{ 1 \over \alpha^D V_D } \ 
{ 1 \over \Gamma( D + 1 ) } \ 
\sum_l \ 
\Big[ \ ( D + l ) ( D + l - 1 ) \ - \ 
l ( l - 1 ) \ \Big] \ 
{ \Gamma( D + l - 1 ) \over \Gamma( l + 1 ) } 
\ 
{ \mu^{ 2 s } \over 
\Big[ 
{ l ( l + D - 1 ) \over \alpha^2 } 
+ m^2 
\Big]^s } 
\eqno(2.12)
$$

Transform the sum over discrete values of $l$ into a 
continuous integral by substituting $l / \alpha \rightarrow k$ 
and $1 / \alpha \rightarrow d k$. We also assume $l \gg 1$, 
yielding $( D + l ) ( D + l - 1 ) \ - \ l ( l - 1 ) 
\rightarrow 2 D l = 2 D k \alpha$, $\Gamma( D + l - 1 ) / 
\Gamma( l + 1 ) \rightarrow l^{ D - 2 }$, and $l ( l + D - 1 ) 
/ \alpha^2 \rightarrow k^2$. This gives 

$$
{ 1 \over \alpha^D V_D }
\zeta( s ) \ = \ 
{ 2 D \ \mu^{ 2 s } \over \Gamma( D + 1 ) \ V_D } 
\int 
{ k^{ D - 1 } \ d k \over ( k^2 + m^2 )^s } 
\ = \ 
{ 2 D \ \mu^{ 2 s } \over \Gamma( D + 1 ) \ V_D } 
{ 1 \over V_{ D - 1 } } 
\int { d^D k \over ( k^2 + m^2 )^s } \ \ \ \ . 
\eqno(2.13)
$$

Now substitute the value $V_D = 2 \pi^{ D + 1 \over 2 } / 
\Gamma( { D + 1 \over 2 } )$ 

$$
{ 1 \over \alpha^D V_D }
\zeta( s ) \ = \ 
{ 
2 D \ \mu^{ 2 s } \ 
\Gamma( { D \over 2 } ) \ \Gamma( { D + 1 \over 2 } ) 
\over 
\Gamma( D + 1 ) \ 4 \pi^{ D + 1 / 2 } 
} 
\int { d^D k \over ( k^2 + m^2 )^s } 
\ = \ 
{ 
\mu^{ 2 s } \ \Gamma( { D + 2 \over 2 } ) \ 
\Gamma( { D + 1 \over 2 } ) 
\over 
\Gamma( D + 1 ) \pi^{ D + 1 / 2 } 
} 
\int { d^D k \over ( k^2 + m^2 )^s } 
\eqno(2.14)
$$

And finally use the identity $\Gamma( z ) 
\Gamma( z + 1 / 2 ) / \Gamma( 2 z ) = 
2^{ 1 - 2 z } \pi^{ 1 / 2 }$ 

$$
{ 1 \over \alpha^D V_D }
\zeta( s ) \ = \ 
{ 
\mu^{ 2 s } \ 2^{ -D } 
\over 
\pi^D 
} 
\int { d^D k \over ( k^2 + m^2 )^s } 
\ = \ 
\int { d^D k \over ( 2 \pi )^D } \ 
{ \mu^{ 2 s } \over ( k^2 + m^2 )^s } \ \ \ \ , 
\eqno(2.15)
$$

which is the Minkowski $\zeta$-density. 

Now back to the de Sitter zeta-function, 
use the transformation [4] 

$$
{ 1 \over 
\big[ \alpha^2 - \beta^2 \big]^\gamma } 
\ = \ 
{ \sqrt{ \pi } \over \Gamma( \gamma ) } 
\int\limits_0^{ \infty } 
\bigg( { t \over 2 \beta } \bigg)^
{ \gamma - 1 / 2 } 
e^{ - \alpha t } \ 
I_{ \gamma - 1 / 2 } \, ( \beta t ) \ dt \ \ \ \ . 
\eqno(2.16)
$$

We get 

$$
\zeta(s) \ = \ 
{ \mu^{ 2 s } \alpha^{ 2s } \over \Gamma( D + 1 ) } \ 
\sum_l \ 
D ( D + 2 l - 1 ) \ \ 
{ \Gamma( D + l - 1 ) \over \Gamma( l + 1 ) } 
\ 
{ \sqrt{ \pi } \over \Gamma( s ) } \ 
\int\limits_0^{ \infty } 
\bigg( { t \over 2 \beta } \bigg)^{ s - 1 / 2 } 
e^{ - ( l + { D - 1 \over 2 } ) t } \ 
I_{ s - 1 / 2 } \, ( \beta t ) \ dt \ \ \ \ , 
\eqno(2.17)
$$

Concentrate on the terms involving $l$, 
and evaluate the sum 

$$
\sum_l \ 
D ( D + 2 l - 1 ) \ 
{ \Gamma( D + l - 1 ) \over \Gamma( l + 1 ) } \ 
e^{ - { 1 \over 2 } ( D + 2 l - 1 ) t } 
\eqno(2.18)
$$

$$
\ = \ 
- 2 D \ { \partial \over \partial t } 
\bigg[ \ 
\sum_l 
{ \Gamma( D + l - 1 ) \over \Gamma( l + 1 ) } \ 
e^{ - { 1 \over 2 } ( D + 2 l - 1 ) t } \ 
\bigg] 
\eqno(2.19)
$$

$$
\ = \ 
- 2 D \ { \partial \over \partial t } 
\bigg[ \ 
e^{ - { D - 1 \over 2 } t } 
\sum_l 
{ \Gamma( D + l - 1 ) \over \Gamma( l + 1 ) } \ 
e^{ - l t } \ 
\bigg] 
\ = \ 
- 2 D \ { \partial \over \partial t } 
\bigg[ \ 
e^{ - { D - 1 \over 2 } t } \ 
{ \Gamma( D - 1 ) \over ( 1 - e^{ - t } )^{ D - 1 } } 
\bigg] 
\eqno(2.20)
$$

$$
\ = \ 
\Gamma( D + 1 ) \ 
{ 2 \cosh( t / 2 ) \over [ 2 \sinh( t / 2 ) ]^D } 
\eqno(2.21)
$$

Now use this result in the expression of the zeta-function 

$$
\zeta( s ) \ = \ 
{ \mu^{ 2 s } \alpha^{ 2s } \over \Gamma( D + 1 ) } \ 
{ \sqrt{ \pi } \over \Gamma( s ) } 
\int\limits_0^{ \infty } 
\bigg( { t \over 2 \beta } \bigg)^{ s - 1 / 2 } \ 
\bigg[ \ 
\Gamma( D + 1 ) \ 
{ 2 \cosh( t / 2 ) \over [ 2 \sinh( t / 2 ) ]^D } \ 
\bigg] \ 
I_{ s - 1 / 2 } \, ( \beta t ) \ dt 
\eqno(2.22)
$$

With an appropriate correction factor, one can extend the contour 
of integration from $- \infty$ to $\infty$, integrating above 
the origin. We will need to introduce a branch cut somewhere 
in the complex plane between the origin and infinity; let this 
branch cut be on the negative imaginary axis (i.e. from the 
origin to $- i \infty$), which corresponds to the contour 
passing the origin on the positive imaginary side to avoid the 
cut. 

$$
\zeta( s ) \ = \ 
{ \mu^{ 2 s } \alpha^{ 2 s } \sqrt{ \pi } \over \Gamma( s ) } \ \ 
{ 1 \over 1 - e^{ i \pi ( 2 s - D ) } } 
\int\limits_{ - \infty + i \Delta } 
^{ \infty + i \Delta } 
\bigg( { t \over 2 \beta } \bigg)^{ s - 1 / 2 } \ 
{ 2 \cosh( t / 2 ) \over [ 2 \sinh( t / 2 ) ]^D } \ 
I_{ s - 1 / 2 } \, ( \beta t ) \ dt 
\eqno(2.23)
$$

Here $\Delta$ is an arbitrary positive quantity, but less than 
$2 \pi$ to stay between the origin and the first pole of the integrand 
on the positive imaginary axis. This expression is the desired 
analytic continuation that is well-defined in a neighborhood of 
$s = 0$. 

\bigskip

{ \bf III. Deforming the contour } 

\smallskip

Use an integral representation of the Bessel-function [5]

$$
\zeta( s ) \ = \ 
{ \mu^{ 2 s } \alpha^{ 2 s } \sqrt{ \pi } \over \Gamma( s ) } \ \ 
{ 1 \over 1 - e^{ i \pi ( 2 s - D ) } } 
\int\limits_{ - \infty + i \Delta } 
^{ \infty + i \Delta } 
\bigg( { t \over 2 \beta } \bigg)^{ s - 1 / 2 } \ 
{ 2 \cosh( t / 2 ) \over [ 2 \sinh( t / 2 ) ]^D } \ 
$$

$$
\bigg[ \ 
{ 1 \over \sqrt{ \pi } \ \Gamma( s ) } 
\bigg( { \beta t \over 2 } \bigg)^{ s - 1 / 2 } \ 
\int\limits_{ -1 }^{ 1 } 
{ e^{ \beta t u } 
\over ( 1 - u^2 )^{ 1 - s } } 
d u \ 
\bigg] \ d t 
\eqno(3.1)
$$

$$
\ = \ 
{ \mu^{ 2 s } \alpha^{ 2 s } \over 
[ \Gamma( s ) ]^2 } \ \ 
{ 1 \over 1 - e^{ i \pi ( 2 s - D ) } } 
\int\limits_{ - \infty + i \Delta } 
^{ \infty + i \Delta } 
\bigg( { t \over 2 } \bigg)^{ 2 s - 1 } \ 
{ 2 \cosh( t / 2 ) \over [ 2 \sinh( t / 2 ) ]^D } \ \ \ 
\int\limits_{ -1 }^{ 1 } 
{ e^{ \beta t u } 
\over ( 1 - u^2 )^{ 1 - s } } \ 
d u \ d t 
\eqno(3.2)
$$

We want to expand this expression in $s$ around $s=0$, so that we can 
deduce $\zeta$ and its first derivative from the result. These are the 
quantities that determine the effective potential. 

First examine the integral over u: 

$$
\int\limits_{ -1 }^{ 1 } 
{ e^{ \beta t u } 
\over ( 1 - u^2 )^{ 1 - s } } \ 
d u 
\ = \ 
\int\limits_{ 0 }^{ 1 } 
{ e^{ \beta t u } 
+ 
e^{ - \beta t u } 
\over ( 1 - u^2 )^{ 1 - s } } \ 
d u 
\ = \ 
\int\limits_{ 0 }^{ 1 } \ 
{ e^{ \beta t ( 1 - v ) } 
+ 
e^{ - \beta t ( 1 - v ) } 
\over v^{ 1 - s } ( 2 - v )^{ 1 - s } } \ 
d v 
\eqno(3.3)
$$

$$
\ = \ 
e^{ \beta t } 
\int\limits_{ 0 }^{ 1 } 
{ e^{ - \beta t v } 
\over v^{ 1 - s } ( 2 - v )^{ 1 - s } } \ 
d v \ \ 
+ \ \ 
( \ \beta t \leftarrow \rightarrow - \beta t \ ) 
\eqno(3.4)
$$

But 

$$
e^{ \beta t } 
\int\limits_{ 0 }^{ 1 } 
{ e^{ - \beta t v } 
\over v^{ 1 - s } ( 2 - v )^{ 1 - s } } \ 
d v 
\ = \ 
e^{ \beta t } \ 
\bigg\{ \ 
\int\limits_{ 0 }^{ 1 } 
{ 1 
\over v^{ 1 - s } ( 2 - v )^{ 1 - s } } \ 
d v \ 
- 
\int\limits_{ 0 }^{ 1 } 
{ e^{ - \beta t v } - 1 
\over v^{ 1 - s } ( 2 - v )^{ 1 - s } } \ 
d v \ 
\bigg\} 
\eqno(3.5)
$$

The first term yields 

$$
{ 1 \over 2 s } + 
\ln 2 + \hbox{O}( s ) 
\eqno(3.6)
$$

The second term can be written as 

$$
\int\limits_{ 0 }^{ 1 } 
{ 1 
\over 
v^{ 1 - s } ( 2 - v )^{ 1 - s } } \ 
\sum_{ n = 1 }^{ \infty } 
{ ( - \beta t v )^n \over n ! } \ \ 
d v 
\eqno(3.7)
$$

$$
\ = \ 
\int\limits_{ 0 }^{ 1 } 
{ 1 
\over 
v ( 2 - v ) } \ \ 
\big[ \ 
1 + s \ln v + \hbox{O}( s^2 ) \ 
\big] \ \ 
\big[ \ 
1 + s \ln ( 2 - v ) + \hbox{O}( s^2 ) \ 
\big] \ 
\sum_{ n = 1 }^{ \infty } 
{ ( - \beta t v )^n \over n ! } \ \ 
d v 
\eqno(3.8)
$$

Since we are looking only for the first two powers of $s$ 
in the expansion, and we have a $1 \ / \ 2 s$ term already, 
we will not need the terms proportional to $s$. Therefore 
we only need 

$$
\int\limits_{ 0 }^{ 1 } 
{ 1 
\over 
v ( 2 - v ) } \ 
\sum_{ n = 1 }^{ \infty } 
{ ( - \beta t v )^n \over n ! } \ 
d v 
\eqno(3.9)
$$

which gives 

$$
\sum_{ n = 1 }^{ \infty } 
{ ( - \beta t )^n \over n ! } \ 
\bigg( \ 
2^{ n - 1 } \ln 2 \ - 
\sum_{ m = 1 }^{ n - 1 } 
{ 2^{ n - m - 1 } \over m } \ 
\bigg) 
\eqno(3.10)
$$

So 

$$
\zeta( s ) 
\ = \ 
{ \mu^{ 2 s } \alpha^{ 2 s } \over 
[ \Gamma( s ) ]^2 } \ \ 
{ 1 \over 1 - e^{ i \pi ( 2 s - D ) } } 
\int\limits_{ - \infty + i \Delta } 
^{ \infty + i \Delta } 
\bigg( { t \over 2 } \bigg)^{ 2 s - 1 } \ 
{ 2 \cosh( t / 2 ) \over [ 2 \sinh( t / 2 ) ]^D } 
$$

$$
\Bigg\{ \ 
e^{ \beta t } \ 
\bigg[ \ 
{ 1 \over 2 s } \ + \ 
\ln 2 \ + \ 
\sum_{ n = 1 }^{ \infty } 
{ ( - \beta t )^n \over n ! } \ 
\bigg( \ 
2^{ n - 1 } \ln 2 \ - 
\sum_{ m = 1 }^{ n - 1 } 
{ 2^{ n - m - 1 } \over m } \ 
\bigg) \ + \ 
\hbox{O}( s ) \ 
\bigg] \ + \ 
( \ \beta t \leftarrow \rightarrow - \beta t \ ) \ 
\Bigg\} 
\eqno(3.11)
$$

Here we have not yet expanded the terms outside the curly 
brackets in powers of $s$, we shall do it later. Evaluate 
the above expression for $m \alpha \gg 1$. This means 
$\beta = \sqrt{ ( D - 1 )^2 / 4 - m^2 \alpha^2 } 
\approx \pm i m \alpha$. 

Let's find a way of closing the contour of integration at 
infinity in the upper or lower half plane, as appropriate for 
the different terms. First analyze the expression 

$$
e^{ i m \alpha t } \ 
\bigg[ \ 
{ 1 \over 2 s } \ + \ 
\ln 2 \ + \ 
\sum_{ n = 1 }^{ \infty } 
{ ( - i m \alpha t )^n \over n ! } \ 
\bigg( \ 
2^{ n - 1 } \ln 2 \ - 
\sum_{ m = 1 }^{ n - 1 } 
{ 2^{ n - m - 1 } \over m } \ 
\bigg) \ + \ 
\hbox{O}( s ) \ 
\bigg] 
\eqno(3.12)
$$

We can write $\ln 2$ as 
$\sum_{ m = 1 }^{ \infty } { 2^{ - m } / m } $, 
and with this, 

$$
\sum_{ n = 1 }^{ \infty } 
{ ( - i m \alpha t )^n \over n ! } \ 
\bigg( \ 
2^{ n - 1 } \ln 2 \ - 
\sum_{ m = 1 }^{ n - 1 } 
{ 2^{ n - m - 1 } \over m } \ 
\bigg) 
\ = \ 
\sum_{ n = 1 }^{ \infty } 
{ ( - i m \alpha t )^n \over n ! } 
\sum_{ m = n }^{ \infty } 
{ 2^{ n - m - 1 } \over m } 
\eqno(3.13)
$$

$$
\ = \ 
\sum_{ n = 1 }^{ \infty } 
{ ( - i m \alpha t )^n \over ( n + 1 ) ! } 
\sum_{ m = 1 }^{ \infty } 
{ 2^{ - m } \over 1 + { m - 2 \over n + 1 } } 
\eqno(3.14)
$$

which clearly shows an $e^{ -i m \alpha t } 
/ ( -i m \alpha t ) $ asymptotic behavior for 
$| m \alpha t | \gg 1$. This combined with the 
$e^{ i m \alpha t }$ in front yields 
$1 / ( -i m \alpha t )$, which cancels with the 
similar term from $\beta t \leftarrow \rightarrow 
- \beta t$, so we can subtract it from the expression. 
Then the remaining expression will go to zero at 
least as fast as $1 / ( m^2 \alpha^2 t^2 )$. 
This allows as to close the contour for this 
term in the upper half plane. The integrand has 
poles in the upper half plane at $2 \pi i n$, 
$n = 1 , 2 ...$ . Applying the residue theorem, 
we obtain the integral as a sum 

$$
{ \mu^{ 2 s } \alpha^{ 2 s } \over 
[ \Gamma( s ) ]^2 } \ \ 
{ 1 \over 1 - e^{ i \pi ( 2 s - D ) } } 
\int\limits_{ - \infty + i \Delta } 
^{ \infty + i \Delta } 
\bigg( { t \over 2 } \bigg)^{ 2 s - 1 } \ 
{ 2 \cosh( t / 2 ) \over [ 2 \sinh( t / 2 ) ]^D } 
$$

\vskip-6pt
\hfill (3.15)
\vskip-18pt

$$
e^{ i m \alpha t } \ 
\bigg[ \ 
- { e^{ -i m \alpha t } 
\over -i m \alpha t } \ + \ 
{ 1 \over 2 s } \ + \ 
\ln 2 \ + \ 
\sum_{ n = 1 }^{ \infty } 
{ ( -i m \alpha t )^n \over n ! } \ 
\bigg( \ 
2^{ n - 1 } \ln 2 \ - 
\sum_{ m = 1 }^{ n - 1 } 
{ 2^{ n - m - 1 } \over m } \ 
\bigg) \ + \ 
\hbox{O}( s ) \ 
\bigg] 
d t 
$$

$$
\ = \ 
{ \mu^{ 2 s } \alpha^{ 2 s } \over 
[ \Gamma( s ) ]^2 } \ \ 
{ 1 \over 1 - e^{ i \pi ( 2 s - D ) } } \ 
\sum_{ n = 1 }^{ \infty } \ 
2 \pi i \ \ { 1 \over \Gamma( D ) } \ 
{ \partial^{ D - 1 } \over 
\partial t^{ D - 1 } } 
\Bigg|_{ t = 2 \pi i n } 
\Bigg\{ \ 
\bigg( { t \over 2 } \bigg)^{ 2 s - 1 } \ 
2 \cosh( t / 2 ) \ 
\bigg( { t - 2 \pi i n \over 
2 \sinh( t / 2 ) } \bigg)^D 
$$

\vskip-6pt
\hfill (3.16)
\vskip-18pt

$$
e^{ i m \alpha t } \
\bigg[ \
- { e^{ -i m \alpha t }
\over -i m \alpha t } \ + \
{ 1 \over 2 s } \ + \
\ln 2 \ + \
\sum_{ n = 1 }^{ \infty }
{ ( -i m \alpha t )^n \over n ! } \
\bigg( \
2^{ n - 1 } \ln 2 \ -
\sum_{ m = 1 }^{ n - 1 }
{ 2^{ n - m - 1 } \over m } \
\bigg) \ + \
\hbox{O}( s ) \
\bigg] \ 
\Bigg\} 
$$

From this form it can be seen that this contribution 
is finite, and it goes to zero with $m \alpha \rightarrow 
\infty$. 

For the $e^{ -i m \alpha t }$ term, we could similarly 
close the contour in the lower half plane, except that 
the integrand has a cut from the origin to $-i \infty$, 
because of the $(t / 2)^{ 2 s - 1 }$. So the curve 
that we use to close the contour must come back from 
$-i \infty$ to $0$ in the positive real half plane, 
then go around the singularity at the origin 
anticlockwise, and go back to $-i \infty$ in the 
negative real half plane, before continuing the 
arc to $-\infty$. The complete contour constructed 
this way does not encircle any singularities, therefore 
the loop integral is zero, which means, that the 
contribution of the piece of contour along the cut 
cancels the contribution of the piece above the 
real axis, which is the integral that we want to 
determine (the two quarter-arcs from $\infty$ to 
$-i \infty$ and from $-i \infty$ to $-\infty$ 
contribute zero, similarly to the semi-arc in the 
upper half plane in the previous case). So our 
integral above the real axis will equal to the 
integral from $-i \infty$ to 
$-i \epsilon$ on the branch continuous with the 
left side of the cut, then a circle around zero 
clockwise with radius $\epsilon$, and then back 
from $-i \epsilon$ to $-i \infty$. 

Now back to the previous expression, let's 
expand it in powers of $s$. 

$$
I( s ) 
\ = \ 
{ \mu^{ 2 s } \alpha^{ 2 s } \over 
[ \Gamma( s ) ]^2 } \ \ 
{ 1 \over 1 - e^{ i \pi ( 2 s - D ) } } 
\int\limits_{ - \infty + i \Delta } 
^{ \infty + i \Delta } 
\bigg( { t \over 2 } \bigg)^{ 2 s - 1 } \ 
{ 2 \cosh( t / 2 ) \over [ 2 \sinh( t / 2 ) ]^D } 
$$

\vskip-6pt
\hfill (3.17)
\vskip-18pt

$$
e^{ -i m \alpha t } \
\bigg[ \
- { e^{ i m \alpha t }
\over i m \alpha t } \ + \
{ 1 \over 2 s } \ + \
\ln 2 \ + \
\sum_{ n = 1 }^{ \infty }
{ ( i m \alpha t )^n \over n ! } \
\bigg( \
2^{ n - 1 } \ln 2 \ -
\sum_{ m = 1 }^{ n - 1 }
{ 2^{ n - m - 1 } \over m } \
\bigg) \ + \
\hbox{O}( s ) \
\bigg]
d t
$$

\vbox{

$$
\ = \ 
{ 1 \over 
-2 \pi i } \ \ 
[ s + s^2 ( 2 \ln ( \mu \alpha ) - 
2 \Gamma'(1) + i \pi ) + 
\hbox{O}( s^3 ) ] 
\int\limits_{ c } 
\bigg( { 2 \over t } [ 1 + 2 s \ln 
{ t \over 2 } + \hbox{O}( s^2 ) ] \ 
{ 2 \cosh( t / 2 ) \over [ 2 \sinh( t / 2 ) ]^D } 
$$

\vskip-6pt
\hfill (3.18)
\vskip-18pt

$$
e^{ -i m \alpha t } \
\bigg[ \
- { e^{ i m \alpha t }
\over i m \alpha t } \ + \
{ 1 \over 2 s } \ + \
\ln 2 \ + \
\sum_{ n = 1 }^{ \infty }
{ ( i m \alpha t )^n \over n ! } \
\bigg( \
2^{ n - 1 } \ln 2 \ -
\sum_{ m = 1 }^{ n - 1 }
{ 2^{ n - m - 1 } \over m } \
\bigg) \ + \
\hbox{O}( s ) \
\bigg]
d t
$$

}

\bigskip

{ \bf IV. Evaluation of the integral }

\smallskip

Let's split the integral into the following terms: 

$$
\eqalign{ 
I_1 &= 
\int\limits_{ c } \ 
{ 2 \over t } \ 
{ 2 \cosh( t / 2 ) \over [ 2 \sinh( t / 2 ) ]^D } \ 
e^{ -i m \alpha t } \ 
{ 1 \over 2 s } \ 
d t \cr \cr
I_2 &= 
\int\limits_{ c } \ 
{ 2 \over t } \ 
2 s \ln { t \over 2 } \ 
{ 2 \cosh( t / 2 ) \over [ 2 \sinh( t / 2 ) ]^D } \ 
e^{ -i m \alpha t } \ 
{ 1 \over 2 s } \ 
d t \cr \cr
I_3 &= 
\int\limits_{ c } \ 
{ 2 \over t } \ 
{ 2 \cosh( t / 2 ) \over [ 2 \sinh( t / 2 ) ]^D } \ 
e^{ -i m \alpha t } \ 
{ -e^{ i m \alpha t } 
\over i m \alpha t } \ 
d t \cr \cr
I_4 &= 
\int\limits_{ c } \ 
{ 2 \over t } \ 
{ 2 \cosh( t / 2 ) \over [ 2 \sinh( t / 2 ) ]^D } \ 
e^{ -i m \alpha t } \ 
\ln 2 \ 
d t \cr \cr
I_5 &= 
\int\limits_{ c } \ 
{ 2 \over t } \ 
{ 2 \cosh( t / 2 ) \over [ 2 \sinh( t / 2 ) ]^D } \ 
e^{ -i m \alpha t } \ 
\sum_{ n = 1 }^{ \infty } 
{ ( i m \alpha t )^n \over n ! } \ 
\bigg( \ 
2^{ n - 1 } \ln 2 \ - 
\sum_{ m = 1 }^{ n - 1 } 
{ 2^{ n - m - 1 } \over m } \ 
\bigg) \ 
d t \cr \cr
}
\eqno(4.1)
$$

keeping only the two lowest-order terms in $s$. 
Only $I_2$ has a cut along the negative imaginary axis, 
so the other terms will decompose into the contributions 
of a series of poles along the negative imaginary axis. 
The contribution from all the poles except the origin can be 
expressed as 

$$
{ \mu^{ 2 s } \alpha^{ 2 s } \over 
[ \Gamma( s ) ]^2 } \ \ 
{ 1 \over 1 - e^{ i \pi ( 2 s - D ) } } \ 
\sum_{ n = 1 }^{ \infty } \ 
2 \pi i \ { 1 \over \Gamma( D ) } \ \ 
{ \partial^{ D - 1 } \over 
\partial t^{ D - 1 } } 
\Bigg|_{ t = -2 \pi i n } 
\Bigg\{ \ 
\bigg( { t \over 2 } \bigg)^{ 2 s - 1 } \ 
2 \cosh( t / 2 ) \ 
\bigg( { t - 2 \pi i n \over 
2 \sinh( t / 2 ) } \bigg)^D 
$$

\vskip-6pt
\hfill (4.2)
\vskip-18pt

$$
e^{ -i m \alpha t } \
\bigg[ \
- { e^{ i m \alpha t }
\over i m \alpha t } \ + \
{ 1 \over 2 s } \ + \
\ln 2 \ + \
\sum_{ n = 1 }^{ \infty }
{ ( i m \alpha t )^n \over n ! } \
\bigg( \
2^{ n - 1 } \ln 2 \ -
\sum_{ m = 1 }^{ n - 1 }
{ 2^{ n - m - 1 } \over m } \
\bigg) \ + \
\hbox{O}( s ) \
\bigg] \ 
\Bigg\} 
$$

which is finite and goes to zero with 
$m \alpha \rightarrow \infty$, similarly to 
the contribution of the poles in the upper-half 
plane. 

Now calculate the conribution from the 
pole at the origin for all but $I_2$. 

$$
\eqalign{
I_1 \ &= \ 
{ 1 \over 2 s } 
\int\limits_{ c } \ 
{ 2 \over t } \ 
{ 2 \cosh( t / 2 ) \over [ 2 \sinh( t / 2 ) ]^D } \ 
e^{ -i m \alpha t } \ 
d t \cr \cr
\ &= \
{ 1 \over 2 s } 
\int\limits_{ c } 
d t \ 
{ 4 \over t^3 } \ 
\bigg[ \ 1 \ + \ t^2 
\bigg( { 1 \over 8 } - 
{ D \over 24 } \bigg) \ + \ 
... \ \bigg] \ 
\bigg( \ 1 \ - \ i m \alpha t \ - \ 
{ 1 \over 2 } m^2 \alpha^2 t^2 \ + \ 
... \ \bigg) \cr \cr
\ &= \ 
{ 2 \over s } \ ( -2 \pi i ) \ 
\bigg( { 1 \over 24 } - { 1 \over 2 } m^2 \alpha^2 \bigg) 
\ = \ 
s^{ -1 } \ 2 \pi i \ ( \ m^2 \alpha^2 - 1 / 12 \ ) \cr \cr
}
\eqno(4.3)
$$

\bigskip

$$
\eqalign{
I_3 \ &= \ 
\int_{ c } \ 
{ 2 \over t } \ 
{ 2 \cosh( t / 2 ) \over [ 2 \sinh( t / 2 ) ]^D } \ 
{ -1 \over i m \alpha t } \ 
d t \cr \cr
\ &= \ 
- { 1 \over i m \alpha } 
\int\limits_{ c } 
d t \ 
{ 4 \over t^4 } \ 
\bigg[ \ 1 \ + \ t^2 
\bigg( { 1 \over 8 } - 
{ D \over 24 } \bigg) \ + \ 
... \ \bigg] 
\ = \ 
0 \cr \cr
}
\mkern220mu
\eqno(4.4)
$$

\bigskip

$$
\eqalign{
I_4 \ &= \ 
\ln 2 
\int\limits_{ c } \ 
{ 2 \over t } \ 
{ 2 \cosh( t / 2 ) \over [ 2 \sinh( t / 2 ) ]^D } \ 
e^{ -i m \alpha t } \ 
d t \cr \cr
\ &= \ 
\ln 2 
\int\limits_{ c } 
d t \ 
{ 4 \over t^3 } \ 
\bigg[ \ 1 \ + \ t^2 
\bigg( { 1 \over 8 } - 
{ D \over 24 } \bigg) \ + \ 
... \ \bigg] \ 
\bigg( \ 1 \ - \ i m \alpha t \ - \ 
{ 1 \over 2 } m^2 \alpha^2 t^2 \ + \ 
... \ \bigg) \cr \cr
\ &= \ 
{ 4 \ln 2 } \ ( -2 \pi i ) \ 
\bigg( { 1 \over 24 } - { 1 \over 2 } m^2 \alpha^2 \bigg) 
\ = \ 
4 \pi i \ \ln 2 \ ( m^2 \alpha^2 - 1 / 12 ) \cr \cr
}
\eqno(4.5)
$$

\bigskip

$$
\eqalign{
I_5 \ &= \ 
\int\limits_{ c } \ 
{ 2 \over t } \ 
{ 2 \cosh( t / 2 ) \over [ 2 \sinh( t / 2 ) ]^D } \ 
e^{ -i m \alpha t } \ 
\sum_{ n = 1 }^{ \infty } 
{ ( i m \alpha t )^n \over n ! } \ 
\bigg( \ 
2^{ n - 1 } \ln 2 \ - 
\sum_{ i = 1 }^{ n - 1 } 
{ 2^{ n - i - 1 } \over i } \ 
\bigg) \ 
d t \cr \cr
\ &= \ 
\int\limits_{ c } 
d t \ 
{ 4 \over t^3 } \ 
\bigg[ \ 1 \ + \ t^2 
\bigg( { 1 \over 8 } - 
{ D \over 24 } \bigg) \ + \ 
... \ \bigg] \ 
\bigg( \ 1 \ - \ i m \alpha t \ - \ 
{ 1 \over 2 } m^2 \alpha^2 t^2 \ + \ 
... \ \bigg) \cr \cr
&\bigg( \ i m \alpha t \ln 2 \ - \ 
{ 1 \over 2 } m^2 \alpha^2 t^2 
( 2 \ln 2 - 1 ) \ - \ 
{ i \over 6 } m^3 \alpha^3 t^3 
\bigg( 4 \ln 2  - { 5 \over 2 } \bigg) \ + \ 
... \ \bigg) \cr \cr
\ &= \ 
4 \ ( -2 \pi i ) \ 
1 / 2 m^2 \alpha^2 
\ = \ 
-4 \pi i \ m^2 \alpha^2
}
\eqno(4.6)
$$

And now calculate $I_2$ on the contour 
that consists of a straight part from 
$-i \infty$ to $-i \epsilon$ on the branch of 
the $\ln$ function continuous with the 
left side of the cut, then a circle around 
the origin with radius $\epsilon$, and 
then another straight part from $-i \epsilon$ 
to $-i \infty$. First calculate the contribution 
of the straight parts. Since the direction of 
integration on the two sides of the cut is opposite, 
only the contribution of the discontinuity across 
the cut remains, which can be expressed as 

$$
{ 1 \over 2 s } 
\int\limits_{ \epsilon }^{ \infty } 
d x \ 
{ 2 \over x } \ 
2 s \ ( -2 \pi i ) \ 
{ 2 \cos( x / 2 ) \over [ -2 i \sin( x / 2 ) ]^D } \ 
e^{ -m \alpha x } 
\eqno(4.7)
$$

$$
= 
\int\limits_{ \epsilon }^{ \infty } 
d x \ 
{ 4 \over x^3 } \ 
( -2 \pi i ) \ 
i^D \ 
\bigg[ \ 1 \ - \ x^2 \ 
\bigg( { 1 \over 8 } - 
{ D \over 24 } \bigg) \ + \ 
... \ \bigg] \ 
e^{ -m \alpha x } 
\eqno(4.8)
$$

$$
= 
\bigg[ \ 
8 \pi i \ 
{ -1 \over 2 x^2 } \ 
e^{ -m \alpha x } \ 
\bigg]_{ \epsilon }^{ \infty } \ 
- \ 
8 \pi i \ 
\int\limits_{ \epsilon }^{ \infty } 
d x \ 
{ -1 \over 2 x^2 } \ 
( -m \alpha ) \ 
e^{ -m \alpha x } \ 
- \ 
8 \pi i \ 
\int\limits_{ \epsilon }^{ \infty } 
d x \ 
{ 1 \over 24 x } \ 
e^{ -m \alpha x } 
\eqno(4.9)
$$

$$
= 
4 \pi i \ 
\epsilon^{ -2 } \ 
e^{ -m \alpha \epsilon } \ 
- \ 
\bigg[ \ 
8 \pi i \ 
{ 1 \over 2 x } \ 
( -m \alpha ) \ 
e^{ -m \alpha x } \ 
\bigg]_{ \epsilon }^{ \infty } \ 
+ \ 
\int\limits_{ \epsilon }^{ \infty } 
d x \ 
4 \pi i \ 
{ 1 \over x } \ 
( m^2 \alpha^2 - 1 / 12 ) \ 
e^{ -m \alpha x } 
\eqno(4.10)
$$

$$
= 
8 \pi i \ 
\epsilon^{ -2 } \ 
e^{ -m \alpha \epsilon } \ 
+ \ 
4 \pi i \ 
\epsilon^{ -1 } 
m \alpha \ 
e^{ -m \alpha \epsilon } 
$$

\vskip-6pt
\hfill (4.11)
\vskip-18pt

$$
+ \ 
\bigg[ \ 
4 \pi i \ 
\ln y \ 
( m^2 \alpha^2 - 1 / 12 ) \ 
e^{ -y } \ 
\bigg]_{ \epsilon m \alpha }^{ \infty } \ 
+ \ 
4 \pi i \ 
\int\limits_{ \epsilon m \alpha }^{ \infty } \ 
d y \ 
\ln y \ 
( m^2 \alpha^2 - 1 / 12 ) \ 
e^{ -y } 
$$

\bigskip

Considering that $\int\limits_{ 0 }^{ \infty } d y \ln y e^{ -y } 
= \Gamma'( 1 )$, we get 

$$
= 
4 \pi i \ 
\epsilon^{ -2 } \ 
( 2 + \epsilon m \alpha ) \ 
e^{ -m \alpha \epsilon } \ 
- \ 
4 \pi i \ 
\ln ( \epsilon m \alpha ) \ 
( m^2 \alpha^2 - 1 / 12 ) \ 
e^{ -\epsilon m \alpha } 
$$

\vskip-6pt
\hfill (4.12)
\vskip-18pt

$$
\ + \ 
4 \pi i \ 
( m^2 \alpha^2 - 1 / 12 ) \ 
\bigg( \ \Gamma'( 1 ) \ - 
\int\limits_{ 0 }^{ \epsilon m \alpha } \ 
d y \ \ln y \ e^{ -y } \bigg) 
$$

Our procedure is to take $\epsilon \rightarrow 0$ 
before we let $m \alpha$ go to infinity. This means 
that we can expand this result in $\epsilon$ and discard all 
terms with positive exponents. 

$$
= 
4 \pi i \ 
\epsilon^{ -2 } \ 
( 2 + \epsilon m \alpha ) \ 
( \ 1 \ - \ \epsilon m \alpha \ + \ 
1 / 2 \ \epsilon^2 m^2 \alpha^2 \ ) \ 
- \ 
4 \pi i \ 
\ln ( \epsilon m \alpha ) \ 
( m^2 \alpha^2 - 1 / 12 ) 
$$

\vskip-6pt
\hfill (4.13)
\vskip-18pt

$$
\ + \ 
4 \pi i \ 
( m^2 \alpha^2 - 1 / 12 ) \ 
\Gamma'( 1 ) 
$$

$$
= 
4 \pi i \epsilon^{ -2 } \ - \ 
8 \pi i \epsilon^{ -1 } m \alpha \ + \ 
6 \pi i m^2 \alpha^2 \ 
- \ 
4 \pi i \ 
( m^2 \alpha^2 - 1 / 12 ) \ 
\ln ( \epsilon m \alpha ) \ 
+ \ 
4 \pi i \ 
( m^2 \alpha^2 - 1 / 12 ) \ 
\Gamma'( 1 ) 
\eqno(4.14)
$$

Now we must calculate the contribution 
of the circle around the origin. 
The radius of the circle is $\epsilon$, 
and $t$ can be expressed as 
$\epsilon e^{ i \theta }$. This means 
that $\ln t = \ln \epsilon + i \theta$, 
$d t$ is equivalent to $i t d \theta$, 
and the integration 
around the circle from $-i \epsilon$ 
clockwise means that $\theta$ runs from 
$3 / 2 \pi$ to $-\pi / 2$. So for this part 
of the integral we have 

$$
{ 1 \over 2 s } 
\int\limits_{ { 3 \over 2 } \pi } 
^{ - { \pi \over 2 } } 
i t \ d \theta \ 
{ 2 \over t } \ 
2 s \ \bigg( \ln { \epsilon \over 2 } 
+ i \theta \bigg) \ 
{ 2 \cosh( t / 2 ) \over [ 2 \sinh( t / 2 ) ]^D } \ 
e^{ -i m \alpha t } 
\eqno(4.15)
$$

$$
= 
2 i \ 
\ln { \epsilon \over 2 } \ 
\int\limits_{ { 3 \over 2 } \pi } 
^{ - { \pi \over 2 } } 
d \theta \ 
{ 2 \cosh( t / 2 ) \over [ 2 \sinh( t / 2 ) ]^D } \ 
e^{ -i m \alpha t } \ 
- \ 
2 
\int\limits_{ { 3 \over 2 } \pi } 
^{ - { \pi \over 2 } } 
\theta \ 
d \theta \ 
{ 2 \cosh( t / 2 ) \over [ 2 \sinh( t / 2 ) ]^D } \ 
e^{ -i m \alpha t } 
\eqno(4.16)
$$

$$
= 
2 i \ 
\ln { \epsilon \over 2 } 
\int\limits_{ { 3 \over 2 } \pi } 
^{ - { \pi \over 2 } } 
d \theta \ 
{ 2 \over t^2 } \ 
\bigg[ \ 1 \ + \ t^2 
\bigg( { 1 \over 8 } - 
{ D \over 24 } \bigg) \ + \ 
... \ \bigg] \ 
\bigg( \ 1 \ - \ i m \alpha t \ - \ 
{ 1 \over 2 } m^2 \alpha^2 t^2 \ + \ 
... \ \bigg) 
$$

\vskip-6pt
\hfill (4.17)
\vskip-18pt

$$
+ 2 
\int\limits_{ - { \pi \over 2 } } 
^{ { 3 \over 2 } \pi } 
\theta \ 
d \theta \ 
{ 2 \over t^2 } \ 
\bigg[ \ 1 \ + \ t^2 
\bigg( { 1 \over 8 } - 
{ D \over 24 } \bigg) \ + \ 
... \ \bigg] \ 
\bigg( \ 1 \ - \ i m \alpha t \ - \ 
{ 1 \over 2 } m^2 \alpha^2 t^2 \ + \ 
... \ \bigg) 
$$

$$
= 
4 i \ 
\ln { \epsilon \over 2 } \ 
( -2 \pi ) \ 
\bigg( { 1 \over 24 } - 
{ 1 \over 2 } m^2 \alpha^2 \bigg) \ 
+ \ 
4 
\int\limits_{ - { \pi \over 2 } } 
^{ { 3 \over 2 } \pi } 
\theta \ 
d \theta \ 
\bigg( \ t^{ -2 } \ - \ 
i m \alpha t^{ -1 } \ + \ 
{ 1 \over 24 } - 
{ 1 \over 2 } m^2 \alpha^2 \ + \ ... \ \bigg) 
\eqno(4.18)
$$

Here we drop the terms that contain $t$ 
to a positive power, because on the circle 
$t = \epsilon e^{ i \theta }$ for $t$, and 
those terms will vanish as 
$\epsilon \rightarrow 0$. Thus we have 

$$
-8 \pi i \ 
\ln { \epsilon \over 2 } \ 
\bigg( { 1 \over 24 } - 
{ 1 \over 2 } m^2 \alpha^2 \bigg) \ 
+ \ 
4 
\int\limits_{ - { \pi \over 2 } } 
^{ { 3 \over 2 } \pi } 
\theta \ 
d \theta \ 
\bigg( \ \epsilon^{ -2 } 
e^{ -2 i \theta } \ - \ 
i m \alpha 
\epsilon^{ -1 } 
e^{ -i \theta } \ + \ 
{ 1 \over 24 } - 
{ 1 \over 2 } m^2 \alpha^2 \ \bigg) 
\eqno(4.19)
$$

$$
= 
-8 \pi i \ 
\ln { \epsilon \over 2 } \ 
\bigg( { 1 \over 24 } - 
{ 1 \over 2 } m^2 \alpha^2 \bigg) \ 
+ \ 
4 \ 
\bigg[ \ 
\theta \ { 1 \over -2 i } 
\epsilon^{ -2 } 
e^{ -2 i \theta } \ - \ 
\theta \ 
i m \alpha \ 
{ 1 \over -i } \ 
\epsilon^{ -1 } 
e^{ -i \theta } \ + \ 
\bigg( { 1 \over 24 } - 
{ 1 \over 2 } m^2 \alpha^2 \bigg) \ 
{ \theta^2 \over 2 } \ 
\bigg]_{ - { \pi \over 2 } } 
^{ { 3 \over 2 } \pi } 
$$

\hfill (4.20)
\vskip-24pt

$$
- 4 
\int\limits_{ - { \pi \over 2 } } 
^{ { 3 \over 2 } \pi } 
d \theta \ 
\bigg( \ { 1 \over -2 i } \ 
\epsilon^{ -2 } 
e^{ -2 i \theta } \ - \ 
i m \alpha \ 
{ 1 \over -i } \ 
\epsilon^{ -1 } 
e^{ -i \theta } \ \bigg) 
$$

$$
= 
-8 \pi i \ 
\ln { \epsilon \over 2 } \ 
\bigg( { 1 \over 24 } - 
{ 1 \over 2 } m^2 \alpha^2 \bigg) \ 
- \ 
4 \pi i \ \epsilon^{ -2 } \ 
+ \ 
8 \pi i \ m \alpha 
\epsilon^{ -1 } \ 
+ \ 
4 \pi^2 \ 
\bigg( { 1 \over 24 } - 
{ 1 \over 2 } m^2 \alpha^2 \bigg) 
\eqno(4.21)
$$

$$
= 
-4 \pi i \ \epsilon^{ -2 } \ + \ 
8 \pi i \ \epsilon^{ -1 } m \alpha \ - \ 
2 \pi^2 \ 
( m^2 \alpha^2 - 1 / 12 ) \ + \ 
4 \pi i \ ( m^2 \alpha^2 - 1 / 12 ) \ 
\ln { \epsilon \over 2 } 
\eqno(4.22)
$$

To obtain $I( s )$, we assemble 
the contribution of the pole at 
the origin and the cut along the negative 
imaginary axis to $\zeta( s )$, 
to first order in $s$. 

$$
I( s ) 
= 
{ \mu^{ 2 s } \alpha^{ 2 s } \over 
[ \Gamma( s ) ]^2 } \ \ 
{ 1 \over 1 - e^{ i \pi ( 2 s - D ) } } \ 
\{ \ 
I_1 + \ 
I_2 + \ 
I_3 + \ 
I_4 + \ 
I_5 \ 
\} 
\eqno(4.23)
$$

\medskip

\vbox{

$$
= 
{ 1 \over 
-2 \pi i } \ \ 
[ \ s \ + \ s^2 ( 2 \ln ( \mu \alpha ) - 
2 \Gamma'( 1 ) + i \pi ) \ + \ 
\hbox{O}( s^3 ) \ ] 
$$

$$
\Bigg\{ \ 
s^{ -1 } \ 2 \pi i \ ( m^2 \alpha^2 - 1 / 12 ) 
$$

$$
+ \ 
4 \pi i \ \epsilon^{ -2 } \ - \ 
8 \pi i \ \epsilon^{ -1 } m \alpha \ + \ 
6 \pi i \ m^2 \alpha^2 \ - \ 
4 \pi i \ 
( m^2 \alpha^2 - 1 / 12 ) \ 
\ln ( \epsilon m \alpha ) \ + \ 
4 \pi i \ 
( m^2 \alpha^2 - 1 / 12 ) \ 
\Gamma'( 1 ) 
$$

$$
- \ 
4 \pi i \ \epsilon^{ -2 } \ + \ 
8 \pi i \ \epsilon^{ -1 } m \alpha \ - \ 
2 \pi^2 \ 
( m^2 \alpha^2 - 1 / 12 ) \ + \ 
4 \pi i \ ( m^2 \alpha^2 - 1 / 12 ) \ 
\ln { \epsilon \over 2 } 
\eqno(4.24)
$$

$$
+ \ 
0 
$$

$$
+ \ 
4 \pi i \ \ln 2 \ ( m^2 \alpha^2 - 1 / 12 ) 
$$

$$
- \ 
4 \pi \ i m^2 \alpha^2 \ 
\Bigg\} 
$$

}

\medskip

$$
= 
{ 1 \over -2 \pi i } \ \
[ \ s \ + \ s^2 ( 2 \ln ( \mu \alpha ) - 
2 \Gamma'( 1 ) + i \pi ) \ + \ 
\hbox{O}( s^3 ) \ ] 
$$

$$
\{ \ 
s^{ -1 } \ 2 \pi i \ ( m^2 \alpha^2 - 1 / 12 ) \ + \ 
2 \pi i \ m^2 \alpha^2 \ - \ 
4 \pi i \ ( m^2 \alpha^2 - 1 / 12 ) \ \ln ( m \alpha ) 
\eqno(4.25)
$$

$$
\ + \ 
4 \pi i \ ( m^2 \alpha^2 - 1 / 12 ) \ \Gamma'( 1 ) \ - \ 
2 \pi^2 \ ( m^2 \alpha^2 - 1 / 12 ) \ 
\} 
$$

\medskip

$$
= 
{ 1 \over -2 \pi i } \ \
\{ \ 
2 \pi i \ ( m^2 \alpha^2 - 1 / 12 ) 
$$

$$
\ + \ 
s \ [ \ 
( 2 \ \ln ( \mu \alpha ) - 2 \Gamma'( 1 ) + i \pi ) \ 
2 \pi i \ ( m^2 \alpha^2 - 1 / 12 ) 
$$

\vskip-6pt
\hfill (4.26)
\vskip-18pt

$$
+ \ 
2 \pi i \ m^2 \alpha^2 \ - \ 
4 \pi i \ ( m^2 \alpha^2 - 1 / 12 ) \ \ln ( m \alpha ) \ + \ 
4 \pi i \ ( m^2 \alpha^2 - 1 / 12 ) \ \Gamma'( 1 ) \ - \ 
2 \pi^2 \ ( m^2 \alpha^2 - 1 / 12 ) \ 
] 
$$

$$
+ \ 
\hbox{O}( s^2 ) \ 
\} 
$$

\medskip

$$
= 
- \ 
( m^2 \alpha^2 - 1 / 12 ) \ + \ 
s \ [ \ -m^2 \alpha^2 \ + \ 
2 ( m^2 \alpha^2 - 1 / 12 ) \ln ( m / \mu ) \ ] \ + \ 
\hbox{O}( s^2 ) 
\eqno(4.27)
$$

From this expression, keeping only the leading behavior 
in $( m \alpha )$, and dividing by $4 \pi \alpha^2$
(the volume of 2-D de Sitter space) to obtain the 
zeta-function density, we find 

$$
\zeta( 0 ) 
\ = \ 
- 
{ m^2 \over 4 \pi }
\eqno(4.28)
$$

and 

$$
\zeta'( 0 ) 
\ = \ 
- 
{ m^2 \over 4 \pi } 
\bigg( 
1 + \ln { \mu^2 \over m^2 } 
\bigg) \ \ \ \ . 
\eqno(4.29)
$$

For the Minkowski zeta-function density in 2-D, we have 
from eq. (1.7) 

$$
\zeta( s )
\ = \
\int
{ d^D k \over ( 2 \pi )^D } \
{ \mu^{ 2 s } \over ( k^2 + m^2 )^s } \ \ \ \ .
\ = \ 
{ m^2 \over 4 \pi }
\bigg(  
{ \mu^2 \over m^2 }
\bigg)^s 
{ 1 \over s - 1 } 
\eqno(4.30)
$$

which gives precisely the values for $\zeta( 0 )$ and $\zeta'( 0 )$ 
that we have obtained above. 

\bigskip

{ \bf V. Conclusions }

\smallskip

In this paper we have, starting with the definition 
of the $\zeta$-function for a minimally coupled 
massive scalar field in D-dimensional de Sitter space, 
given an integral representation for it which is 
well-defined near $s = 0$. We have then deformed 
the contour of integration so as to expose those 
terms which give the leading behavior as the 
de Sitter radius $\alpha$ tends to infinity. 

In this limit, de Sitter space becomes Minkowski 
space, and we verify, specializing to the case 
$D = 2$ for simplicity, that $\zeta( 0 )$ and 
$\zeta'( 0 )$ tend to the expected Minkowski 
values. This is a property that is not at all 
evident from the original integral representation, 
and, as pointed out in the introduction, is a 
property that might have been lost in the process 
of analytic continuation. It is both reassuring 
and a non-trivial check of our computations that 
the Minkowski limit is in fact recovered after 
analytic continuation. 

With these results in hand, one can consider 
the addition of a symmetry-breaking potential 
for the scalar field. One will then be able 
to study questions related to field quantization 
in de Sitter space in the presence of interactions 
that, in Minkowski space, would lead to symmetry 
breaking effects. In particular, one will be 
able to study the way in which particular 
time slicings may lead to symmetry restoration 
whereas others may allow the symmetry to remain 
broken [1]. Investigations along these lines are in 
progress. 

\bigskip

\eject

{ \bf Acknowledgements }

\medskip

Research supported in part by DOE grant no.
DE-AC02-ERO3075.

\bigskip

{ \bf References }

\bigskip

[1] B. Ratra, Phys. Rev. D {\bf 31} , 1931 (1985) 

\ \ \ \ B. Ratra, Phys. Rev. D {\bf 50} , 5252 (1994) 

\ \ \ \ A. Kaiser and A. Chodos, Phys. Rev. D {\bf 53} , 787 (1996). 

\smallskip

[2] J. S. Dowker and R. Critchley, Phys. Rev. D {\bf 13} , 3224 (1976) 

\ \ \ \ S. W. Hawking, Commun. Math. Phys. {\bf 55} , 133 (1977) 

\smallskip

\ \ \ \ E. Elizalde, S. D. Odintsov, A. Romeo, A. A. Bytsenk and S. Zerbini,

\ \ \ \ {\it Zeta Regularization Techniques with Applications},

\ \ \ \ World Scientific Publishing Co. Pre. Ltd., 1994.

\smallskip

[3] A. Chodos and E. Myers, Ann. Phys. {\bf 156} , 412 (1984) 

\smallskip

[4] A. Erd\'elyi, W. Magnus, F. Oberhettinger and F. G. Tricomi, 

\ \ \ \ {\it Higher Transcendental Functions - Bateman manuscript project}, 

\ \ \ \ McGraw-Hill Book Company Inc., 1953. 

\smallskip

[5] M. Abramowitz and I. A. Stegun, {\it Handbook of
mathematical functions},

\ \ \ \ \ Dover books on advanced mathematics, 1972.

\end